\documentclass{article}

\usepackage{arxiv}

\usepackage[utf8]{inputenc} 
\usepackage[T1]{fontenc}    
\usepackage{hyperref}       
\usepackage{url}            
\usepackage{booktabs}       
\usepackage{amsfonts}       
\usepackage{nicefrac}       
\usepackage{microtype}      
\usepackage{lipsum}
\usepackage{amsmath,amssymb,amsfonts}
\usepackage{algorithmic}
\usepackage{graphicx}
\usepackage{textcomp}
\usepackage{xcolor}

{}
{}
{}

\title{Pinched Hysteresis Loops In  Nonlinear Resonators\thanks{The paper has been accepted at IET Circuits, Devices and Systems}}

\author{A. S. Elwakil$^{1,2,3}$, M. E. Fouda $^{4,5}\thanks{Email. foudam@uci.edu}$, S. Majzoub $^{1}$, A. G. Radwan $^{4,3}$ \\
{$^{1}$Department of Electrical and Computer Engineering, University of Sharjah, Sharjah, Emirates}\\
{$^{2}$Department of Electrical and Computer Engineering, University of Calgary, Canada}\\
{$^{3}$Nanoelectronics Integrated Systems Center (NISC), Nile University, Egypt}\\
{$^{4}$Engineering Mathematics and Physics Dept., Cairo University, Egypt}\\
{$^{5}$Electrical Engineering and Computer Science, University of California-Irvine, Irvine, USA}
}

\begin{document}
\maketitle

\begin{abstract}
This paper shows that pinched hysteresis can be observed in  simple nonlinear resonance circuits containing a single diode that behaves as a voltage-controlled switch. Mathematical models are derived and numerically validated for both series and parallel resonator circuits. The lobe area of the pinched loop is found to increase with increased frequency and multiple pinch-points are possible with an odd symmetrical nonlinearity such as a cubic nonlinearity. Experiments have been performed to prove the existence of pinched hysteresis with a single diode and with two anti-parallel diodes. The formation of a pinched loop in these circuits confirms that:
1) pinched hystersis is not a finger-print of memristors and that
2) the existence of a nonlinearity is essential for generating this behavior. Finally, an application in a digital logic circuit is validated.
\end{abstract}


\section{Introduction}

Memristors have been proposed as new electronic devices with promising analog/digital and neuromorphic applications 
\cite{shin2010memristor,zidan2014compensated,fouda2019spiking,wu2019enhanced,karimi2019novel,fouda2014memristor}. Their characteristic behavior is a pinched hysteresis loop in the current-voltage plane \cite{ascoli2015nonlinear}. Many circuits (known as emulators) have been introduced to mimic the memristor's behavior  \cite{ranjan2018high,petrovic2019tunable,elwakil2013simple,alharbi2017electrical,fouda2014simple}. Although this behavior has been shown to exist in other nonlinear devices such as a non-linear inductor or a non-linear capacitor \cite{fouda2015pinched,corinto2012memristive,sadecki2019analysis}, and that its appearance is linked to satisfying the necessary conditions of the theory of Lissajous figures  \cite{maundy2019correlation, maundy2018simple, weilnhammer2016perceptual}, it is still mistakenly attributed  to memristors only. There are actually doubts about the uniqueness of the memristor as a fundamental device which have been raised by several researchers (see \cite{abraham2018case} and the references therein) but there is no doubt that it is a dynamic and nonlinear device \cite{riaza2014comment,elwakil2013simple, non}. Without being nonlinear, the frequency-doubling mechanism mandated by the theory of Lissajous figures, which is essential to create a pinched loop (as explained in detail in \cite{maundy2019correlation} and most recently in \cite{MAJZOUB}) cannot be obtained. It is important to note that even if the memristor was not a fundamental device, it may still be a useful nonlinear device. Therefore, semi-conductor devices labeled as memristors continue to be fabricated in different technologies and materials \cite{babacan2018fabrication, bio} and are now commercially available from some vendors (see for example www.knowm.com). 

Any memristive system is generally defined as 
\cite{chua1976memristive,radwan2015mathematical}

\begin{align}
y=g(x,u,t)u \nonumber\\
\dot{x}=f(x,u,t) \nonumber    
\end{align}
where $u$ and $y$ are the input and output of the system and $x$ is the state variable. To identify a device as a memristor, it has to have three fingerprints 
\cite{adhikari2013three,radwan2015mathematical}
: 1) a pinched loop  in current-voltage plane with
2)  a loop area that decreases monotonically with increasing frequency and 3) shrink to a single-valued function when the frequency tends to infinity. This  does not mean that pinched hysteresis loops cannot be observed in other devices or circuits (with increasing or decreasing loop area).

In this paper, we provide  two examples of  simple fundamental circuits, namely  nonlinear series and parallel resonators, that can exhibit the pinched hysteresis loop behavior. Mathematical models are derived and numerically validated with a diode-type asymmetric switching nonlinearity and also with an odd-symmetric nonlinearity formed of two anti-parallel diodes. Experimental results are  provided and confirm the theory. The formation of a pinched loop from these circuits confirms our previous assertions \cite{fouda2015pinched,maundy2019correlation} that:
\begin{enumerate}
\item pinched hystersis is not a finger-print of memristors only,
\item  pinched loops with lobe area widening (rather than declining) with increased frequency are possible and
\item the existence of a nonlinearity is essential for generating this behavior. This is because the theory of Lissajous figures necessitates the generation of a second harmonic frequency in the electrical current $i(t)$ when the applied voltage on the device $v(t)$ has only one fundamental frequency (see \cite{MAJZOUB} for a more detailed explanation). 
\end{enumerate}

We further demonstrate the use of the generated pinched loop behavior in realizing digital logic AND/OR gates. With this, we aim to show that research in the possible applications of pinched loops \cite{app1, app2} can be conducted without restricting this behavior to memristors.

\section{Nonlinear resonators}

\subsection{Series RLC Resonator}

\begin{figure}[h]
\centering{\includegraphics[width=60mm]{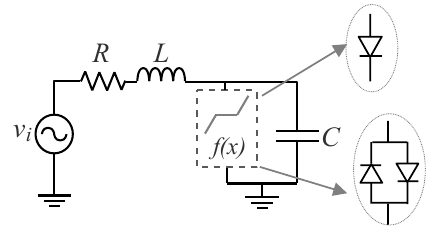}}
\caption{Nonlinear series resonator circuit with diode-based nonlinear switches.}
\label{fig:sercit}
\end{figure}

A simple nonlinear resonator circuit is shown in Fig. \ref{fig:sercit} composed of a series RLC resonance circuit interrupted by a voltage-controlled switching nonlinearity (denoted $f(x)$) placed across the capacitor. The circuit is excited by a sinusoidal signal $v_{in}(t)=V_{DC}+V_A\sin(\omega t)$. In its simplest form, the switching device can be a single diode and in this case the circuit is described by the equation set

\begin{subequations}
\begin{align}
C\frac{dv_{C}}{dt}=i_{L}-i_{f},\\
L\frac{di_{L}}{dt}=v_{in} -Ri_{L}-v_{C}
\end{align}
\end{subequations}
where $v_C$ and $i_L$ are respectively the voltage and current across the capacitor and in the inductor while $i_f$ is the nonlinear device current given by 

\begin{figure}[h]
 \centering{\includegraphics[width=130mm]{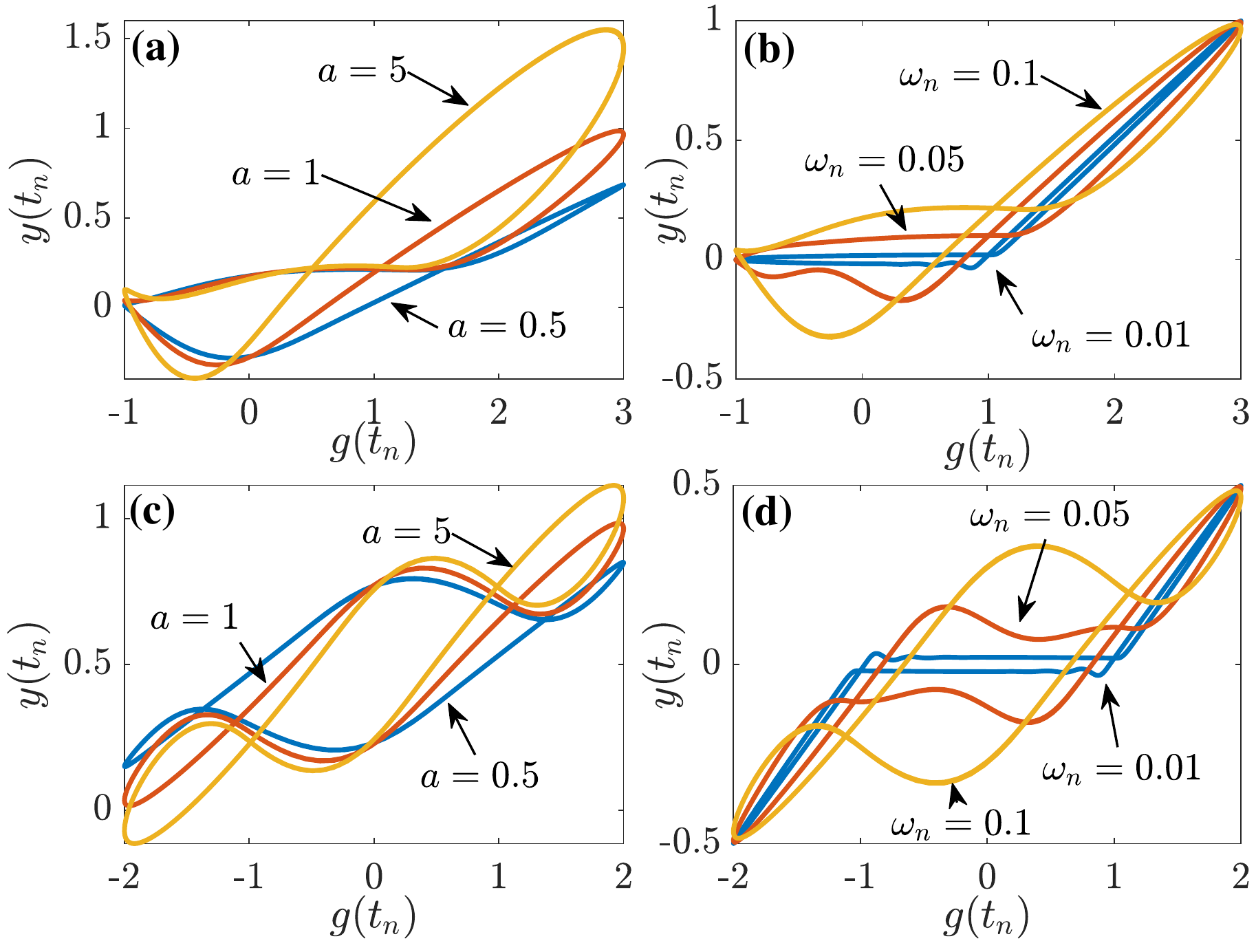}}
 \caption{Numerical simulation results of the nonlinear model of the series resonator circuit with (a), (b) a single diode-nonlinearity  and (c), (d) with two anti-parallel diodes.  The effect of  different values of parameter $a$ at $\omega_n=0.1$ are shown in  (a) and (c) and the effect of different values of $\omega_n$ at $a=1$ are shown in  (b) and (d)}
\label{fig:serNum}
\end{figure}

\begin{equation}
i_f=i_{D}=\left\{ \begin{array}{cc}
(v_{C}-v_{\gamma})/R_{D} & \:\:v_{C}\geq v_{\gamma}\\
0 & \:\:v_{C}<v_{\gamma}
\end{array}\right.
\end{equation}
where $R_D$ is the diode switch-on resistance while $v_{\gamma}$ is the switch-on voltage (approximately $0.6 V$ for silicon devices). Introducing the dimensionless variables: $x=v_{C}/v_{\gamma},y=Ri_{L}/v_{\gamma},a=R/R_{D},b=R^2C/L, A=V_{DC}/v_{\gamma}, B=V_A/v_{\gamma}$, a normalized frequency $\omega_n=\omega RC$ and a normalized time $t_n=t/RC$, the above equations transform into
\begin{subequations}
\begin{align}
\frac{dx}{dt_n}=y-f(x),\\
\frac{dy}{dt_n}=b\left(g(t_n )-y-x\right)
\end{align}
\end{subequations}
with $g(t_n)=A+B\sin(\omega_n t_n)$  being the sinusoidal excitation and $f(x)$ being the nonlinearity. In the case of the switching-diode, $f(x)$ is asymmetric and  given by 

\begin{equation}
f(x)=\left\{ \begin{array}{cc}
a(x-1) & \;x\geq1\\
0 & \;x<1
\end{array}\right.
\end{equation}
Numerical simulations of this model were performed using the ode45 solver in MATLAB with $A=1,b=0.2,B=2$ for different values of $a$ and $\omega_n$ as shown in Figs. \ref{fig:serNum}(a) and 2(b) which plot the observed pinched loop in the $g(t_n)-y(t_n)$ plane; i.e.  in the $v_i-i_L$ plane of the circuit. Therefore, this pinched-loop represents an impedance analogous to the "memristance". It is noted from these simulations that the lobe area increases with increasing the value of the parameter $a$  and also with increasing the frequency $\omega_n$. Note that $\omega_n$ can also be written as $\omega_n=\omega\cdot BW/\omega_{r}^{2}$ where $\omega_r=1/\sqrt{LC}$ is the resonance frequency and $BW=R/L$ is the resonance bandwidth. Hence, satisfying a certain value of $\omega_n$ is achieved not only by changing the applied excitation signal frequency $\omega$ but also by adjusting the resonance bandwidth and resonance frequency.

Based on the detailed analysis and explanation of \cite{maundy2019correlation}, the appearance of the pinched loop in this nonlinear resonator is a result of satisfying the conditions of the theory of Lissajous figures. In particular, the nonlinearity $f(x)$ is responsible for generating a strong second-harmonic  in the current signal $y(t)$ with a normalized frequency $2\omega_n$. The existence of this harmonic can easily be verified from the  FFT of $y(t)$. So long as the phase shift between this harmonic and the fundamental frequency component remains less than $\pi/2$,  the pinched-loop will be observable \cite{maundy2019correlation}. The location of the pinch point can be found  using the general equations derived in \cite{maundy2019correlation}.  When the diode-type nonlinearity is replaced with an odd-symmetric one, such as a cubic-type nonlinearity  or the nonlinearity resulting from two anti-parallel diodes given by (5) (see appendix), two pinched-points are obtained in this case as shown in Figs. \ref{fig:serNum}(c) and (d)  for the parameter values $A=0,b=0.2$  and $B=2$.A pure sinusoid  is used in this case for excitation ($A=0$).

\begin{equation}
f(x)=\left\{ \begin{array}{cc}
a(x-1) & \;x\geq1\\
0 & \;-1<x<1\\
a(x+1) & \;x\leq -1
\end{array}\right. 
\end{equation} 

Shaping of the pinched-loop is thus clearly related to the choice of the  function $f(x)$. In particular, the odd-symmetric nonlinearity generates a third-order harmonic from the applied fundamental frequency of the excitation voltage. The total number of pinch points in a Lissajous figure formed from two sinusoids with frequencies $p \cdot \omega$ and $q \cdot \omega$ is known to equal $q(p-1)+p(q-1)$ which is equal to 2 when $p=1$ and $q=3$.

\subsection{ Parallel RLC Resonator}

\begin{figure}[h]
\centering{\includegraphics[width=45mm]{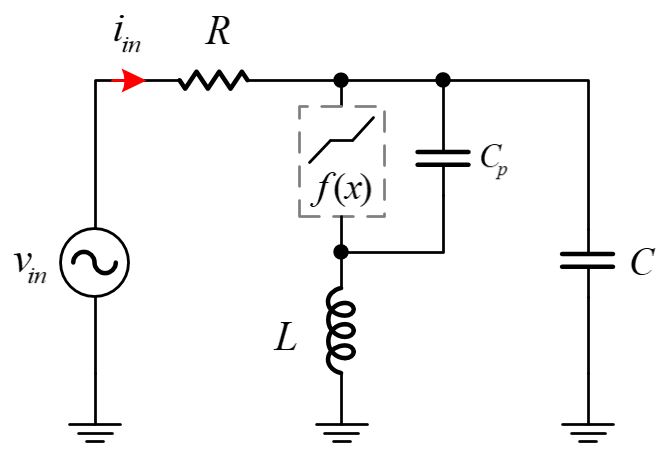}}
\caption{Nonlinear parallel resonator circuit}
\label{fig:parcit}
\end{figure}

Another  nonlinear resonator circuit is shown in Fig. \ref{fig:parcit} composed of a parallel RLC resonance circuit interrupted by nonlinearity placed in series  with the inductor. $C_p$ is a parasitic capacitor that must be included in the analysis since the non-linearity is voltage-controlled. In this case the circuit is described by the equation set

\begin{subequations}
\begin{align}
RC\frac{dv_{C}}{dt}=v_{in}-v_C-i_L R, \\
L \frac{di_L}{dt}=v_c-v_{c_p},\\
C_p \frac{v_{c_p}}{dt}=i_L-i_f
 \end{align}
\end{subequations}

With the same dimensionless variables introduced in the previous subsection, the above equations can be written as

\begin{subequations}
\begin{align}
\frac{dx}{dt_n}=g(t_n )-x-y, \\
\frac{dy}{dt_n}={b}(x-z)\\
\epsilon \frac{dz}{dt_n}=y-f(z)
\end{align}
\end{subequations}
where $z=i_D R/v_\gamma$, $\epsilon=C_p/C$ and $f(z)$ is the switching nonlinearity function given by (4) or (5). When $\epsilon$ tends to zero, (7c) yields $y=f(z)$.

\begin{figure}[h]
\centering{\includegraphics[width=0.8\linewidth]{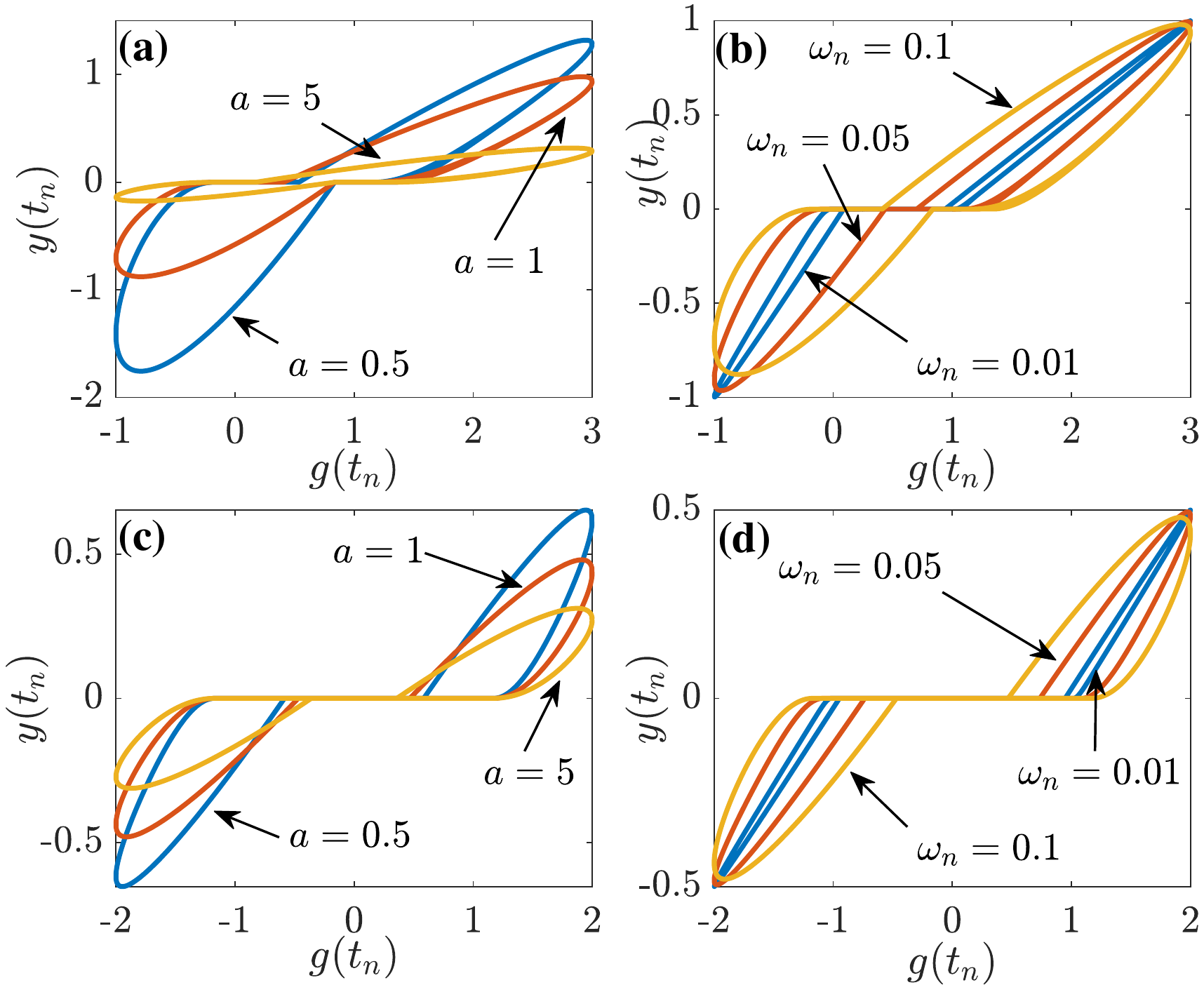}}
\caption{Numerical simulation results of the nonlinear model of the parallel resonator circuit with a single diode-nonlinearity and with two anti-parallel diodes: (a), (c) at $\omega_n=0.1$ for different values of $a$ and (b), (d) {at $a=1$ } for different values of $\omega_n$}
\label{fig:parNum}
\end{figure}

Figure \ref{fig:parNum} shows the numerical simulation of (7) with $A=1,b=0.2,B=2$  and $\epsilon=0.01$. The area enclosed inside the lobes increases with increasing the frequency $\omega_n$ but decreases with increasing the value of parameter $a$.  

\section{Experimental results}
\begin{figure}[h]
\centering{\includegraphics[width=70mm]{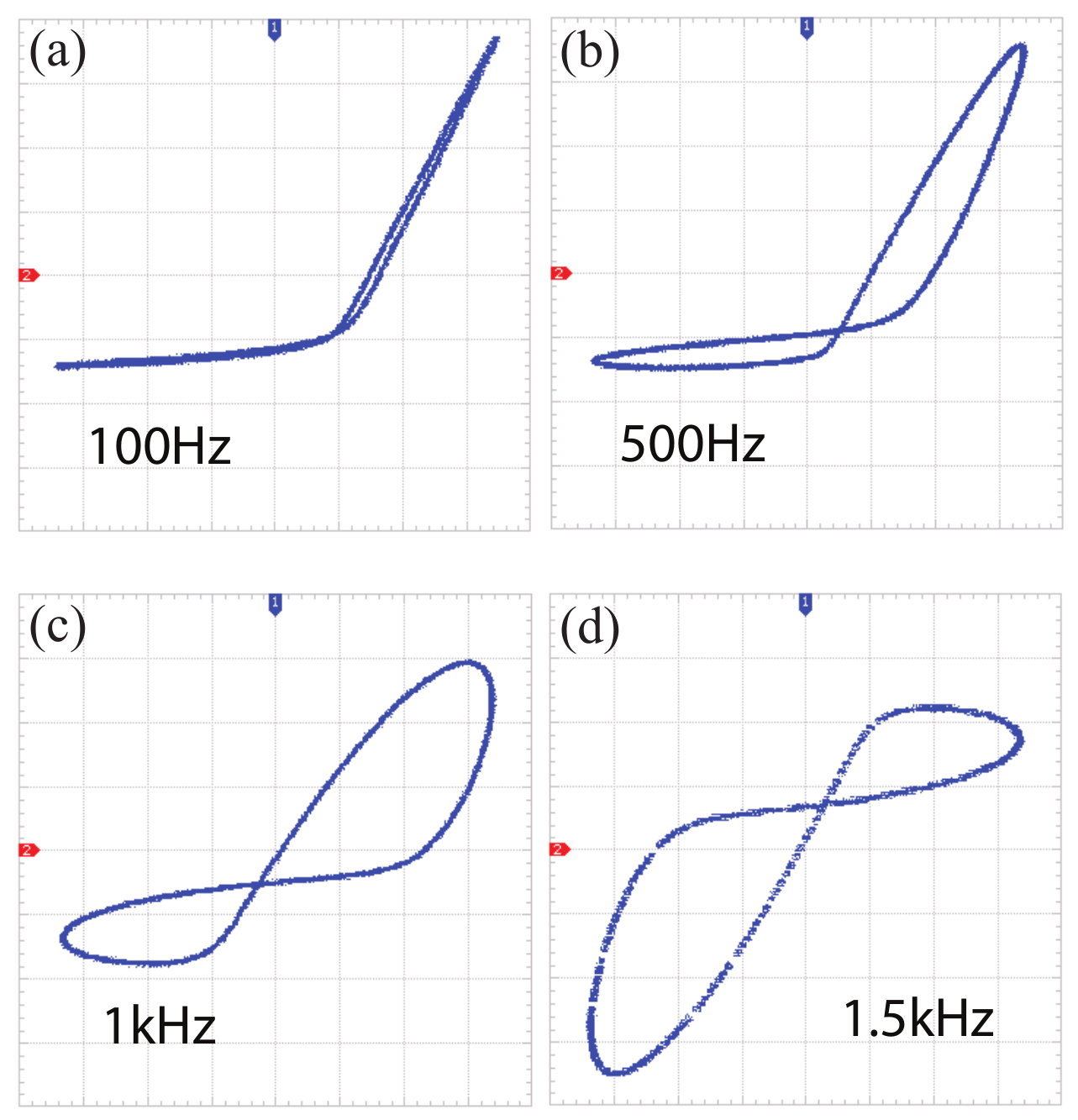}}
\caption{Results from an experimental setup: (a)-(d) formation of the pinched loop at different frequencies (Xaxis $v_{i}(t)$: 0.6V/div, Yaxis: $i_{L}(t)$: 0.3V/div).}
\label{fig:serexp}
\end{figure}

\begin{figure}[h]
\centering{\includegraphics[width=60mm]{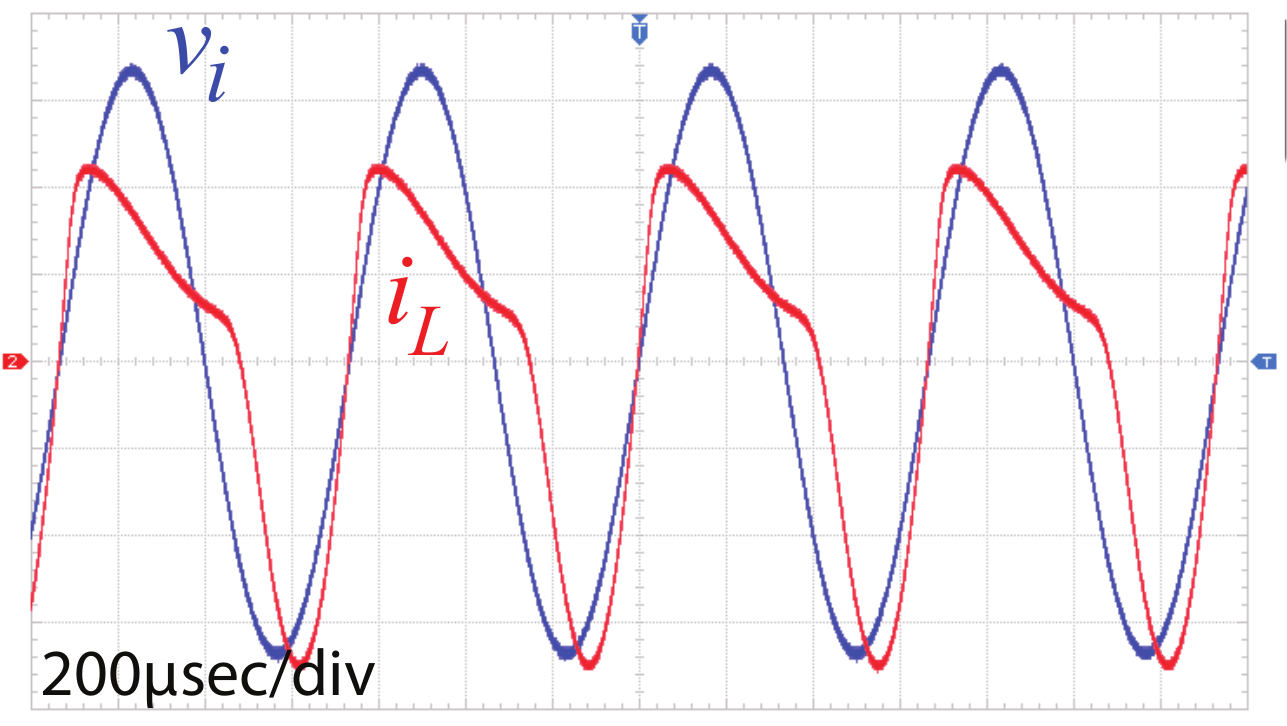}}
\caption{Time domain waveforms at $1.5kHz$}
\label{fig:serexptime}
\end{figure}

\begin{figure}[h]
\centering{\includegraphics[width=85mm]{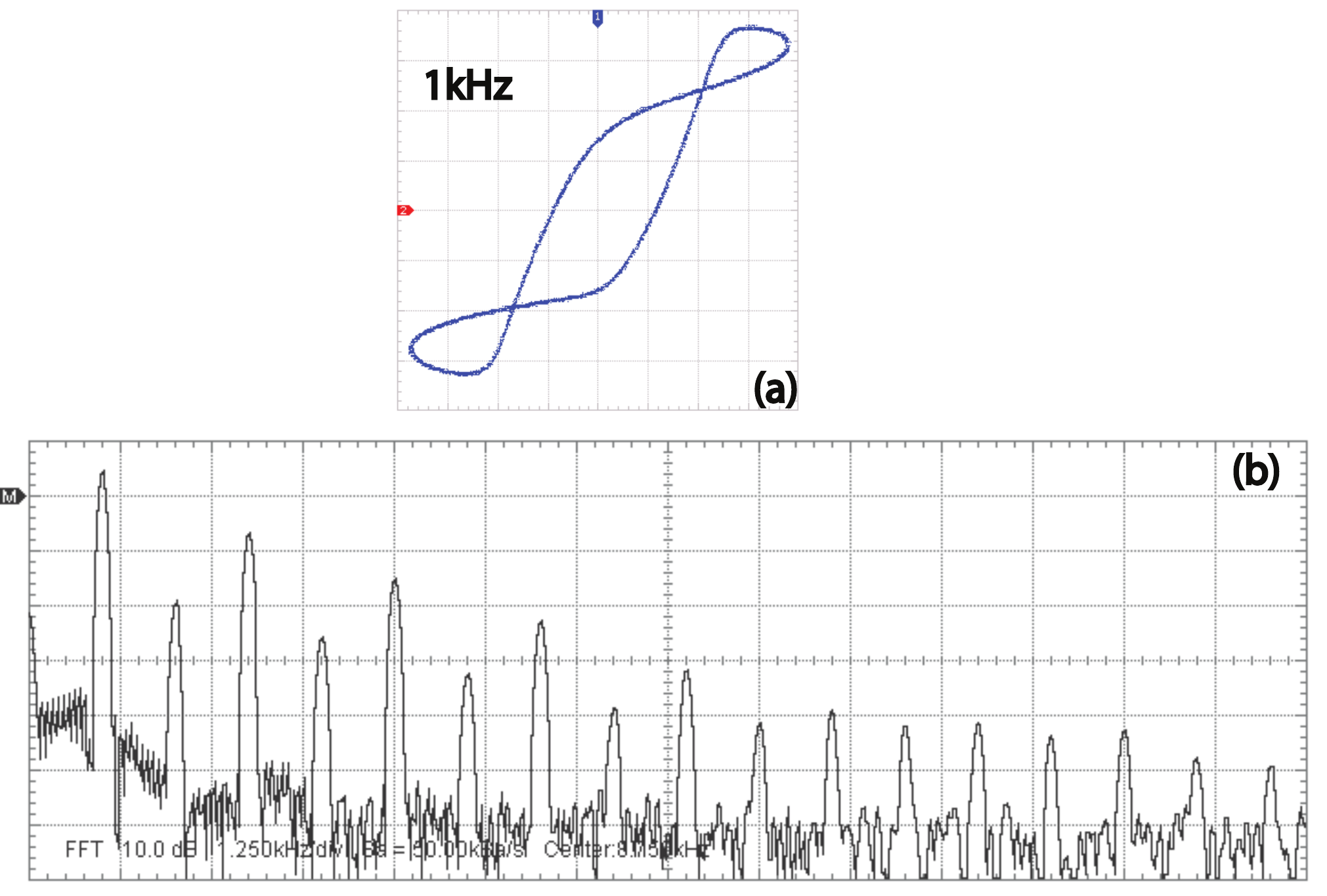}}
\caption{Experimental results using anti-parallel diodes: (a) loop with two pinch points  (Xaxis $v_{i}(t)$: 2V/div, Yaxis: $i_{L}(t)$: 0.65V/div). (b) power spectrum of $i_L(t)$ (Xaxis: 1.25kHz/div, Yaxis: 10dB/div)}
\label{fig:serexpfft}
\end{figure}

The nonlinear series resonator circuit was experimentally tested with discrete components $L=1mH$, $C=5.6 nF$, $R=10k \Omega$ and using a 1N914 diode. Figure \ref{fig:serexp} shows the observed pinched loop when the excitation voltage has an offset voltage $V_{DC}=1V$, an amplitude of $V_A=2V$ and at different frequencies. The current in the resistor was measured using a differential probe.\\
At $100Hz$ (see Fig. \ref{fig:serexp}(a)) it is seen that the $v_i-i_L$ portrait resembles the diode characteristics as expected. As frequency increases, the pinched loop formation is observed in Figs. \ref{fig:serexp}(b), (c) and (d). The voltage and current  waveforms corresponding to the portrait of Fig. \ref{fig:serexp}(d) are shown in Fig. \ref{fig:serexptime}. Adding another diode with opposite polarity in parallel to the diode in the circuit, a loop with two pinch points is seen in Fig. \ref{fig:serexpfft}(a) as predicted numerically. The applied voltage in this case  had no offset value ($V_{DC}=0V$), an amplitude of $V_A=7V$ and a $1kHz$ frequency. The corresponding power spectrum of the current signal $i_L$ is shown in Fig. \ref{fig:serexpfft}(b) from which it is visible that strong odd harmonics have been generated. However, the two pinched points are created because the phase angle between the fundamental  signal at $1kHz$ and the generated third-order harmonic at $3kHz$ is less than $\pi/2$ satisfying the Lissajous conditions \cite{maundy2019correlation}. The remaining harmonics contribute to the shape of the loop but not to the creation of the pinch-points.

\section{Application and Comparison with a Commercial Memristor Device}
\begin{figure}[h]
\centering{\includegraphics[width=70mm]{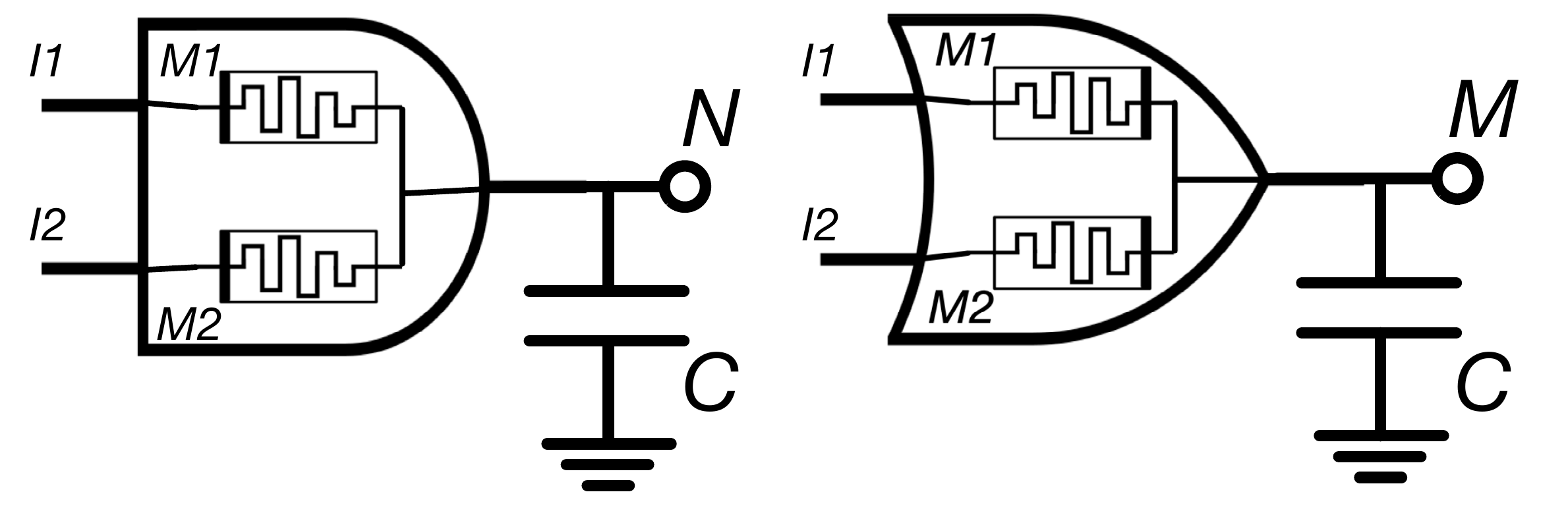}}
\caption{Realization of AND/OR gates using two memristors}
\label{fig:OrAND}
\end{figure}

\begin{figure}[h]
\centering{\includegraphics[width=80mm]{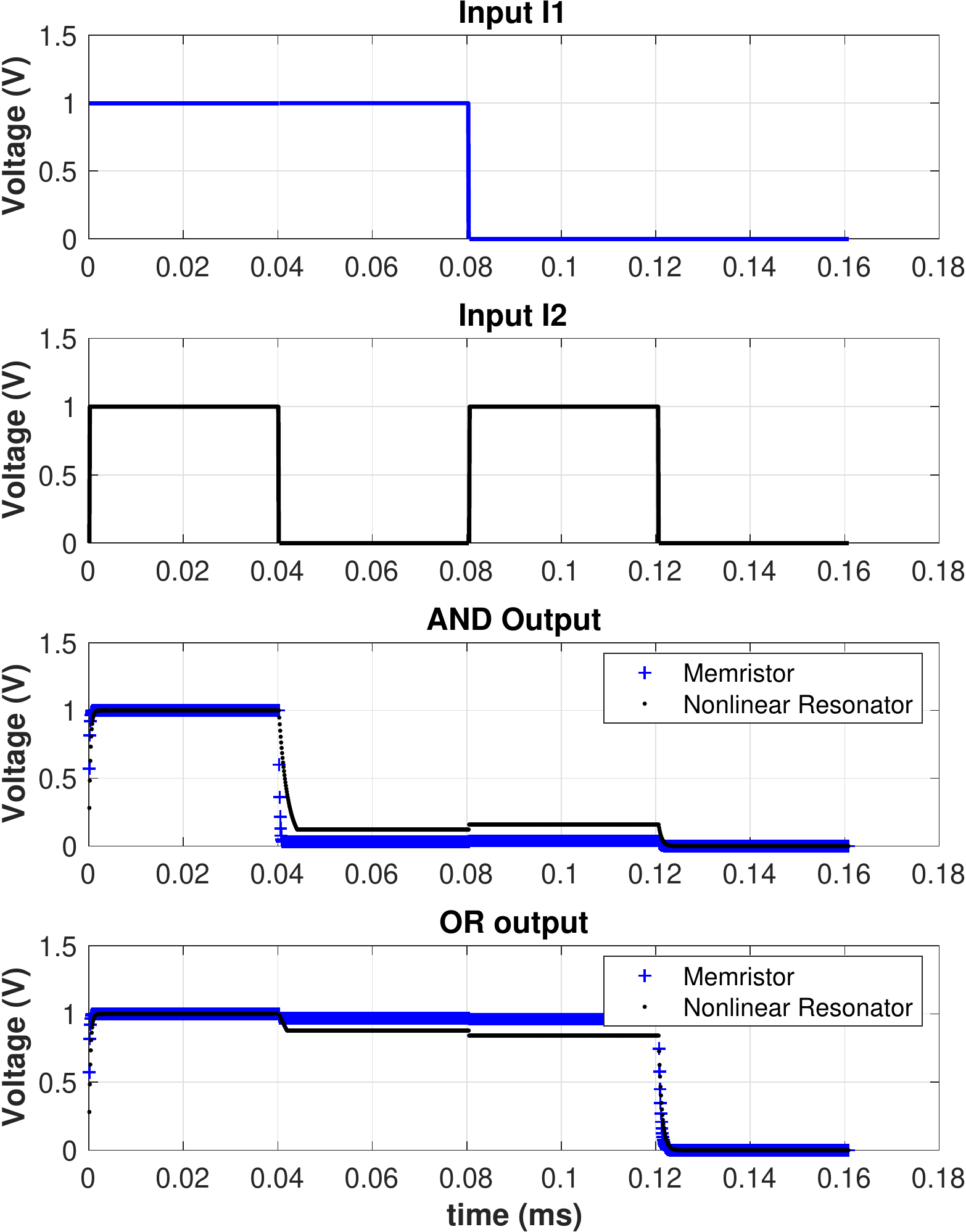}}
\caption{Comparison of the AND/OR gate responses using the pinched loop data from the nonlinear series resonator circuit versus the data from a commercial memristor}
\label{fig:OrANDexp}
\end{figure}

Digital application of AND/OR gates are used to compare their functionality with pinched loops created (i) from the series resonator described above and (ii) from a KNOWM commercial memristor device using their discovery kit \cite{knowm}. The gates are built using two memristor devices, as shown in Fig. \ref{fig:OrAND}, driving a capacitive load (C = 10nF).  Figure \ref{fig:OrANDexp} shows the waveform on the capacitor for both the AND/OR gates after using the experimental memristance data of the pinched loop obtained from the nonlinear resonator circuit and that obtained from  a commercial memristor device \cite{knowm}. It is clear from the figure that similar responses are obtained in both cases. Therefore, from an application point of view, what matters is the existence of the pinched loop rather than its origin. On the other hand, resonators are easily realizable on the micro and nano scales \cite{res1, res2} which paves the way for possibly easier implementations of memristors based on the technique presented here. In terms of the non-volatility of these circuits, a procedure to dis-connect the ground terminal connection should be devised in order to maintain the states of the state variables (inductor current and capacitor voltage) un-changed. It is important to mention, that the non-volatility of memristors is not related to  whether they are fundamental  devices or not.

\section{Conclusion}
We have verified numerically and experimentally the existence of a dynamic pinched-loop behavior in a simple nonlinear resonator circuits. The key behind generating such a loop is to produce in the electrical current a second-order harmonic (for a single pinch point case) or third-order harmonic (for a two pinch point case) from the applied fundamental frequency of the excitation voltage. When these harmonics have proper phase and amplitude relations with respect to the fundamental \cite{maundy2019correlation}, pinched loops appear. The circuits presented here rely on the  diode nonlinearity as opposed to MOS transistor quadratic nonlinearity \cite{maundy2018simple, babacan2018fabrication} or multiplier-type nonlinearity \cite{elwakil2013simple, complexity}. The proposed systems represent  second-order and third-order dynamical systems which is not the case of  memrisitve systems as defined in \cite{chua1976memristive}.

\appendix
\section*{Appendix: Anti-parallel connected diodes}

The total current of the anti-parallel-connected diodes is
\begin{equation}
i_f=i_{D1}+i_{D2}     
\end{equation}
Assuming that $D1$ and $D_2$ are forward and backward connected, respectively, then

\begin{align}
i_{D1}=\left\{ \begin{array}{cc}
(v_{D}-v_{\gamma})/R_{D} & \:\:v_{D}\geq v_{\gamma}\\
0 & \:\:v_{D}<v_{\gamma}
\end{array}\right.\\
i_{D2}=\left\{ \begin{array}{cc}
(v_{D}+v_{\gamma})/R_{D} & \:\:v_{D}\leq -v_{\gamma}\\
0 & \:\:v_{D}>-v_{\gamma}
\end{array}\right.
\end{align}
thus,
\begin{equation}
i_f=\left\{ \begin{array}{cc}
(v_{D}-v_{\gamma})/R_{D} & \:\:v_{D}\geq v_{\gamma}\\
(v_{D}+v_{\gamma})/R_{D} & \:\:v_{D}\leq -v_{\gamma}\\
0 & \:\:-v_{\gamma}<v_{D}<v_{\gamma}
\end{array}\right.     
\end{equation}
which results in eqn. (5) .

\bibliographystyle{unsrt}

\end{document}